# Dielectric Screening by 2D Substrates

*Keian Noori[†,§], Nicholas Lin Quan Cheng[‡,§], Fengyuan Xuan[†], Su Ying Quek[\*,†,‡]*

[†]Centre for Advanced 2D Materials, National University of Singapore, 6 Science Drive 2, 117546, Singapore

[‡]Department of Physics, National University of Singapore, 2 Science Drive 3, 117542, Singapore

KEYWORDS: dielectric screening, 2D materials, energy-level alignment, organic-inorganic interfaces

Abstract

Two-dimensional (2D) materials are increasingly being used as active components in nanoscale devices. Many interesting properties of 2D materials stem from the reduced and highly non-local electronic screening in two dimensions. While electronic screening within 2D materials has been studied extensively, the question still remains of how 2D substrates screen charge perturbations or electronic excitations adjacent to them. Thickness-dependent dielectric screening properties have recently been studied using electrostatic force microscopy (EFM) experiments. However, it was suggested that some of the thickness-dependent trends were due to extrinsic effects. Similarly, Kelvin probe measurements (KPM) indicate that charge fluctuations are reduced when BN slabs are placed on $SiO_2$, but it is unclear if this effect is due to intrinsic screening from BN. In this work, we use first principles calculations to study the fully non-local dielectric screening properties of 2D material substrates. Our simulations give results in good qualitative agreement with those from EFM experiments, for hexagonal boron nitride (BN), graphene and $MoS_2$, indicating that the experimentally observed thickness-dependent screening effects are intrinsic to the 2D materials. We further investigate explicitly the role of BN in lowering charge potential fluctuations arising from charge impurities on an underlying $SiO_2$ substrate, as observed in the KPM experiments. 2D material substrates can also dramatically change the HOMO-LUMO gaps of adsorbates, especially for small molecules, such as benzene. We propose a reliable and very quick method to predict the HOMO-LUMO gap of small physisorbed molecules on 2D and 3D substrates, using only the band gap of the substrate and the gas phase gap of the molecule.

Two-dimensional (2D) materials, being atomically thin and flexible, are prime candidates as active components in next generation nanoscale devices. Many properties of 2D materials, such as charge density wave instabilities,[1] the presence of strongly bound excitons,[2] as well as tunable interlayer excitons,[3-5] arise from the reduced electronic screening in two dimensions. In fact, electronic screening in 2D systems is non-trivial. Unlike in 3D, where screening can be described by a macroscopic dielectric constant, dielectric screening in 2D is highly nonlocal.[6-8] While electronic screening within 2D materials has been studied extensively, the question still remains of how 2D substrates screen charge perturbations or electronic excitations adjacent to them. This question becomes particularly pertinent with recent interest in incorporating 2D materials as part of larger heterostructures.

One of the major breakthroughs in graphene electronics was the discovery that placing graphene on hexagonal boron nitride (BN) substrates on top of $SiO_2$ results in superior electron and hole mobilities as compared with graphene placed directly on $SiO_2$.[9] Subsequent studies have shown that graphene placed on BN (and BN itself) has significantly lower charge density fluctuations than graphene on $SiO_2$.[10, 11] However, it is unclear if the effect results from the migration of charge impurities away from the interface due to the presence of BN, or if it is due simply to a greater spatial separation between graphene and the charge impurities on $SiO_2$, or if BN itself plays a significant role in screening away the effect of the charge impurities. Electrostatic force microscopy (EFM) experiments on three common 2D materials (graphene, $MoS_2$ and BN) placed on $SiO_2$ suggest that electronic screening by atomically thin BN and $MoS_2$ has relatively weak layer number dependence compared to few layer graphene.[12-14] However, the interpretation of these experimental results is complicated by the presence of impurities and charge transfer from the environment, with some of the thickness-dependent trends being explained by the presence of water molecules.[12] Furthermore, the estimated density of defects in the experiment on BN was lower than that on $MoS_2$ and graphene (BN: $10^8$ cm$^{-2}$; $MoS_2$, graphene: $10^{12}$ cm$^{-2}$).[12-14] Taken together, the importance of screening by 2D materials and the questions raised in the experiments underscore the need for first principles studies on the thickness-dependent intrinsic screening properties of layered 2D materials.

Besides screening of unwanted charge impurities, electronic screening by substrates can also change the quasiparticle (QP) gaps of adjacent materials.[15, 16] This is particularly noticeable for small molecular adsorbates. To understand the effect of screening on the QP gap, consider the energy required to remove an electron from the highest occupied molecular orbital (HOMO) to vacuum. Electronic screening from the substrate will stabilize the resultant hole, moving the HOMO level up. Similarly, screening moves the lowest unoccupied molecular orbital (LUMO) level down. These effects are not present in typical density functional theory (DFT) calculations but are captured by many-body GW calculations.[17] The high cost of GW calculations has motivated simpler models to estimate screening from bulk substrates using a classical image charge model.[15, 16] However, screening by 2D substrates is more subtle, due to the difficulty in defining a macroscopic dielectric constant, and the atomically thin nature of the material. Given the growing interest in organic-2D material heterostructures,[18-20] it is timely to explore methods to estimate the energy level alignment at organic-2D material interfaces to facilitate bottom-up design.

In this work, we employ first principles calculations to study the electronic screening of point charge perturbations adjacent to a 2D material. Specifically, we compute *ab initio* $V_{scr}(r,r')$, the screening potential at $r$ due to an electron point charge perturbation at $r'$.[21] The latter represents point charges on an underlying $SiO_2$ substrate, on which a 2D material is placed, or represents the quasiparticle in an adsorbed molecule. We thus simulate *ab initio* the surface potentials of 2D materials in the presence of underlying point charges, and find that our results compare well with data from EFM experiments[12-14] as well as Kelvin probe[11] experiments. We also provide a reliable back-of-the-envelope way to estimate the HOMO-LUMO gap for small adsorbed molecules on 2D and 3D substrates, from the band gap of the substrate and the gas phase gap of the molecule.

We define the screening potential to be $V_{scr} = W - V$, with $V$ being the bare Coulomb potential and $W$ the static screened Coulomb potential. In real space, the screening potential $V_{scr}(r,r')$, with $r'$ representing the position of an extra electron point charge and $r$ representing the position of the probe, is given by the formula[22]

$$V_{scr}(r,r') = \sum_{q,G,G'} e^{i(q+G).r} \left[ \varepsilon^{-1}_{G,G'}(q,\omega=0) - \delta_{G,G'} \right] \times \frac{4\pi e^2}{\Omega |q+G'|^2} e^{-i(q+G').r'}, \qquad (1)$$

where $\varepsilon_{G,G'}(q,\omega=0)$ is the static dielectric function computed using the random phase approximation (RPA). Details of the calculations are given in the Methods section.

Impurity screening by 2D substrates has recently been studied using electrostatic force microscopy (EFM) experiments,[12-14] which effectively measure the difference in surface potentials between the top and bottom of the 2D slabs,[14] in the presence of charged impurities on the underlying $SiO_2$ substrate. Since the difference in surface potentials in non-polar slabs is caused by the induced charge in response to the underlying charged impurity, we can compute this quantity for non-polar slabs of 2D materials using

$$\Delta V_{scr} = V_{scr}(r_{top},r') - V_{scr}(r_{bottom},r'), \qquad (2)$$

where $r'$ is taken to be ~3 Å beneath the bottom atomic plane (approximating the position of charge impurities on an $SiO_2$ substrate), and $r_{top}$ and $r_{bottom}$ are at the top and bottom atomic planes, respectively. We estimate $\Delta V_{scr}$ by taking $r_{top}$ and $r_{bottom}$ to be directly above $r'$, and averaging over the in-plane coordinates of the perturbing charge, $r'$. In Figure 1, we plot the quantity,

$$\Delta V_{scr}^{diff} = |\Delta V_{scr} - \Delta V_{scr}^{bulk}|, \qquad (3)$$

which is the magnitude of the difference between $\Delta V_{scr}$ computed for the thin film and that for the bulk, as a function of film thickness. This is the quantity that is plotted in experiment (its sign varies depending on the sign of the point charges and the direction of charge transfer), and in Figure 1 we have extracted its value for BN[12], MoS$_2$[14], and graphene[13] for comparison with our calculated results. For $\Delta V_{scr}^{bulk}$, we use the value of $\Delta V_{scr}$ computed for the thickest film we have. We find that the qualitative agreement between the computed and experimental values is remarkably good, despite the fact that the values from experiment are influenced by charge transfer effects from the environment.[23] This good agreement implies that the qualitative thickness-dependent screening effects observed in experiment can be explained by intrinsic screening properties of the 2D materials.

In Ref. 12, the experimental results were explained using a non-linear Thomas-Fermi model that took into account interlayer coupling. According to this model, the measured value of $\Delta V_{scr}^{diff}$ for 1-layer BN would be an outlier, and this was attributed to the presence of much more charge transfer between an underlying dipole water film and the 1-layer BN.[12] However, our *ab initio* results indicate that the value of $\Delta V_{scr}^{diff}$ for 1-layer BN fits well with the rest of the data for few-layer BN and the trend observed in experiment agrees well with the trend in our calculations (Figure 1b). We therefore propose that the data for 1-layer BN is not an outlier, and suggest that a non-linear Thomas-Fermi model may not be sufficient for explaining all the non-trivial screening characteristics in 2D substrates.

The different dielectric screening from the environment and differing defect densities in the different experiments[12-14] make it difficult to deduce from experiment the relative intrinsic thickness-dependent screening properties of the various materials. Using our *ab initio* data, we see that $\Delta V_{scr}^{diff}$ changes with thickness most quickly for few-layer graphene, and least quickly for BN. However, the differences we predict are much less drastic than those observed in the experiments.[12-14]

We address the quantitative difference between our *ab initio* results and the experimental data by including the effects of environmental screening using a dielectric constant $\varepsilon$ (Figure 1). The value of $\varepsilon$ is deduced by fitting our *ab initio* results in (3) to the experimental data.

$$\Delta V_{scr}^{model}(1L) = \frac{1}{\varepsilon}\left(V_{scr}^{ab-initio}(1L) - V_{scr}^{ab-initio}(bulk)\right) = V^{exp}(1L) - V^{exp}(bulk) \quad (4)$$

$$\Delta V_{scr}^{model}(n_{max}L) = \frac{1}{\varepsilon}\left(V_{scr}^{ab-initio}(n_{max}L) - V_{scr}^{ab-initio}(bulk)\right) = V^{exp}(n_{max}L) - V^{exp}(bulk) \quad (5)$$

The model results compare very well with the experimental data (Figure 1). We note that our definition of $\Delta V_{scr}$ is likely to overestimate the difference in surface potentials between the top and bottom of the slabs, for typical defect densities, and therefore, our fitted values of $\varepsilon$ should be taken to be an upper bound to the actual value of $\varepsilon$ in the experiment. Nonetheless, we comment that the value of $\varepsilon$ found for the BN data (56.16) is consistent with the presence of water,[24] which is believed to exist in the BN experiment.[12] On the other hand, the value of $\varepsilon$ found for the graphene data (2.32) agrees well with the dielectric constant estimated in Ref. 13.

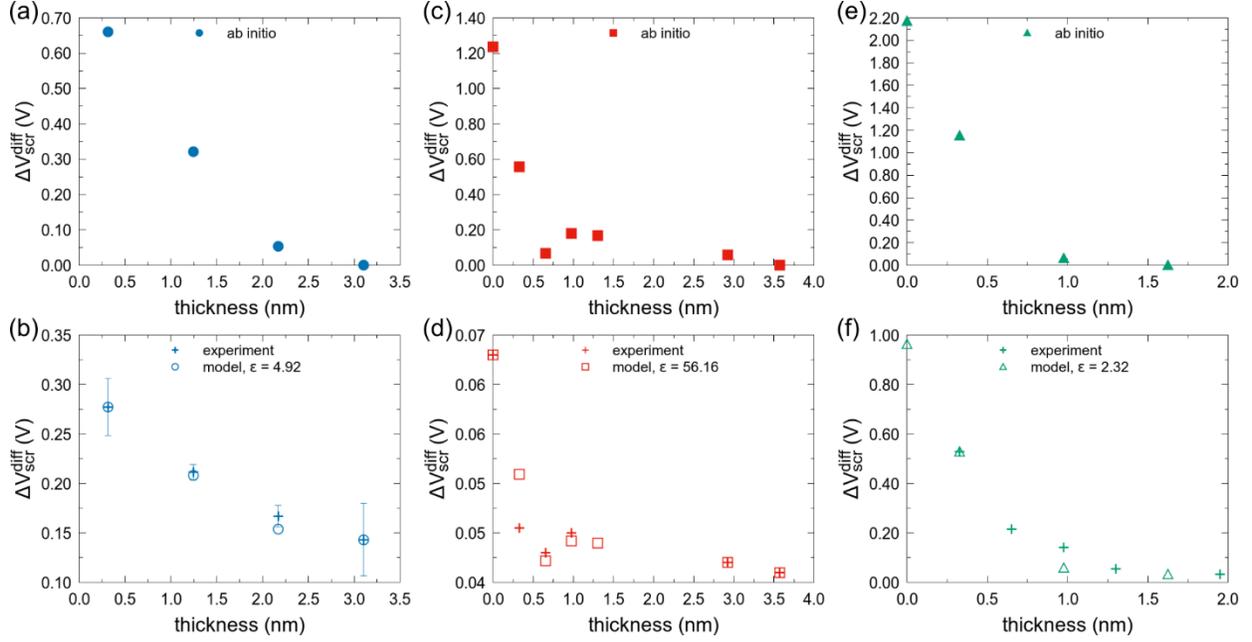

**Figure 1.** Plot of $\Delta V_{scr}^{diff}$ versus layer number for few-layer (a) MoS$_2$, (b) BN and (c) graphene. $\Delta V_{scr}^{diff}$ is defined in the text (equation (3)). The top panels (a, c, e) show computed *ab initio* values. The bottom panels (b, d, f) show experimental values extracted from Refs. 12-14, together with the values obtained by including a model background dielectric constant $\varepsilon$ in the *ab initio* results (using equations (4) and (5)).

It is well known that graphene devices placed atop BN substrates on SiO$_2$ have much higher mobility than graphene devices placed directly atop SiO$_2$.[9] This has been partially attributed to the atomically flat nature of BN. At the same time, experiments also provide evidence of reduced charge potential fluctuations on top of BN, compared to SiO$_2$.[10, 11] For example, in Ref. 11, the autocorrelation function, *C(r)*, computed for the measured surface potential using Kelvin probe microscopy, was smaller in magnitude on top of BN compared to SiO$_2$, from which it was deduced that BN reduced charge potential fluctuations. It is unclear if this effect results from the migration of charge impurities away from the interface due to the presence of BN, or if it is due simply to a greater spatial separation between graphene and the charge impurities on SiO$_2$, or if BN itself plays a significant role in screening away the effect of the charge impurities. We directly address the role of BN in reducing charge potential fluctuations by computing the 2D autocorrelation function, *C(r)*, of the screened and bare Coulomb potentials in different planes above a point charge perturbation, which is taken to be ~3 Å below BN, similar to what is expected for charge impurities on an underlying SiO$_2$ substrate. In order to simulate *C(r)* for large enough *r*, we have to obtain the value of $W(r,r')$ for large $|r-r'|$, which implies that we need a very dense k-point mesh for computing $\varepsilon_{G,G'}(q,\omega=0)$. In this work, we have used a k-point mesh of 73 x 73 x 1 for BN when computing $\varepsilon_{G,G'}(q,\omega=0)$ (corresponding equivalently to a real-space supercell of 73 x 73 x 1 in the expansion of $W(r,r')$), and limited our study to monolayers. From Figure 2a, we can see that *C(r)* for the screened Coulomb potential *W* in a plane 1 Å above the BN surface is smaller, and drops off more slowly with *r*, than *C(r)* for the bare Coulomb potential *V* in a plane 1 Å above the impurity (below the BN layer). These results for *C(r)* are in qualitative agreement with the experimental results in Ref. 11, where the BN slab is thicker. Furthermore, both the rate at which *C(r)* drops off with *r* and the magnitude of *C(r)* are smaller for *W* than for *V* in the same plane above BN. These features imply that intrinsic screening from BN is important for reducing the potential fluctuations created by underlying charge impurities. In Figure 2b, we show *C(r)* for *W* and *V* above 1L BN, 1L MoS$_2$, and graphene, with the supercell sizes of the latter two chosen to closely match the area of the BN supercell. We see that, while *C(r)* for *V* above BN and graphene are the same (the small inconsistency in the plot is due to the slightly different areas of the cells), *C(r)* for *W* of graphene is both flatter and greatly reduced in magnitude compared to BN. This observation is consistent with the larger screening expected from semi-metallic graphene, compared to the large band gap insulator, BN. The smaller magnitude of *C(r)* for *V* above MoS$_2$ compared to BN, meanwhile, is indicative of the greater thickness of the MoS$_2$ layer. Nevertheless, $C(r)$ for $W$ above MoS$_2$ shows potential fluctuation reductions beyond those of the corresponding $V$ curve, demonstrating that MoS$_2$ also effectively screens the charge impurity. Thus, we can conclude that the non-trivial dielectric screening is a general feature of 2D materials, and that the thickness of the screening substrate is important. However, the effective screening of impurities by BN in particular, together with the fact that BN itself is relatively free of dangling bonds and charge impurities, makes it an ideal substrate to improve charge mobility in graphene.[9-11]

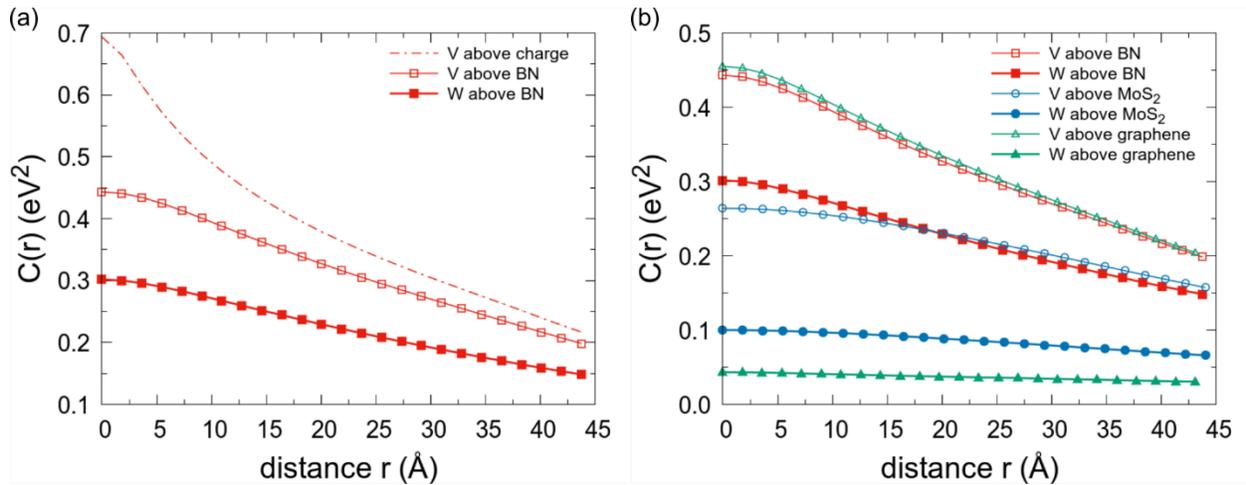

**Figure 2.** The 2D autocorrelation, $C(r)$, for screened ($W$) and bare ($V$) Coulomb potential maps in (a) 1L BN computed for a 73x73x1 supercell, corresponding to a charge impurity density of 3.47 x 10$^{11}$ cm$^{-2}$; and (b) 1L BN, MoS$_2$ and graphene, with supercells chosen to match the area of the BN supercell

Interestingly, the quantity $V_{scr}(r,r')$ is also useful for estimating the HOMO-LUMO gap of small molecules adsorbed on substrates. In particular, the renormalization of molecular levels for molecules physisorbed on a substrate is given by the static polarization integral,[15]

$$P_j = -\frac{1}{2}\iint drdr'\phi_j(r)\phi_j^*(r')\Delta W(r,r')\phi_j(r')\phi_j^*(r), \tag{6}$$

where $\phi_j$ represents the *j*th molecular orbital, and $\Delta W$ is the change in screened Coulomb potential in the molecule upon adsorption. Thus,

$$E_{gap}^{V_{scr}} = E_{gap}^{gas-phase} - (P_{HOMO} + P_{LUMO}) \tag{7}$$

For small molecules, we may approximate $\Delta W(r,r')$ by $\Delta W(r,r) \approx \Delta V_{scr}(r,r)$.[15] It has been shown that for benzene physisorbed on bulk metallic or semiconducting substrates, $\Delta V_{scr}(r,r)$ can be approximated by a classical image potential model. Such an approximation assumes that $\Delta V_{scr}(r,r)$ is given by the screening potential of the substrate alone, and gives good agreement with benchmark GW results for the HOMO-LUMO gap of benzene on various bulk substrates.[15,16] For 2D materials, the classical image potential model cannot be directly applied because of the ambiguity in defining a macroscopic dielectric constant, as illustrated above. By approximating $\Delta V_{scr}(r,r)$ as the screening potential $V_{scr}(r,r)$ for the 2D material substrate, with *r* being the center of the molecule (point charge model), we can estimate the HOMO-LUMO gap of benzene physisorbed on 2D materials without having to do a full GW calculation for the interface, using the following equation:

$$E_{gap}^{V_{scr}} = E_{gap}^{gas-phase} - V_{scr}(r,r), \tag{8}$$

with *r* in the center of the molecule. We use the above method to predict the HOMO-LUMO ($\pi$-$\pi^*$) gaps for single benzene molecules adsorbed on 1- and 2-layer BN, 1-layer $MoS_2$, and graphene. We find that the benzene molecules adsorb at a height of ~3.25 Å above the top atomic plane and use this adsorption height in the calculations. We note that in this regime, small amounts of hybridization between the benzene LUMO and the electronic states of the substrate are present, for BN and $MoS_2$ (Table S1). This is in contrast to previous literature utilizing equation (7), where there is no hybridization between benzene and the substrate, either because of the substrate material or the chosen adsorption heights.[15, 16] Nevertheless, the benzene molecules do not interact strongly with the substrate and it is instructive to see how equations (7-8) fare in these more realistic adsorption geometries. We compare the predictions of equations (7-8) with benchmark GW calculations of the same system in 3 x 3 supercells (as well as a 4x4 supercell for 1L BN) (Figure 3). Notably, the 2D materials lead to significant renormalization of the benzene HOMO-LUMO gap from its gas phase value (10.6 eV). The agreement between $E_{gap}^{V_{scr}}$ and the GW values is reasonable and shows that $V_{scr}(r,r)$ captures most of the physics of screening by the 2D substrates, despite the small amount of hybridization present. We note that the effects of hybridization can be included in our estimate of the HOMO-LUMO gap using more involved methodologies.[25, 26] We can further improve our estimation in equation (8) by using Bader charges[27], centered at each atom of the molecule, instead of using a point charge model. These results are given in Table 1.

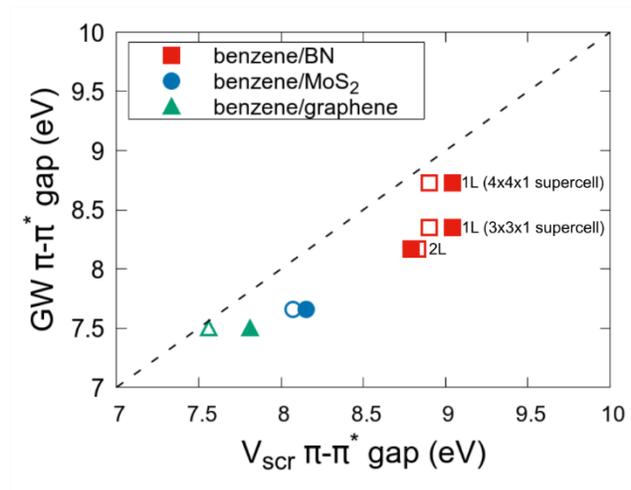

**Figure 3.** GW π-π* gap of benzene on BN, MoS$_2$ and graphene versus the π-π* gap predicted using $V_{scr.}$ (equation (8)). The filled symbols denote the values predicted using the *ab initio* $V_{scr}$

(with a point charge model), while the hollow symbols denote values predicted using $V_{scr}$ obtained from the best fit line in Figure 4.

|  | GW π-π* gap of full interface system (eV) | $E_{gap}^{V_{scr}}$ (eV) |
|---|---|---|
| benzene/1L BN (3x3) | 8.35 | 9.04 [9.21] |
| benzene/1L BN (4x4) | 8.73 | 9.04 [9.18] |
| benzene/2L BN (3x3) | 8.17 | 8.79 [9.08] |
| benzene/1L MoS$_2$ (3x3) | 7.66 | 8.14 [8.36] |
| benzene/graphene (3x3) | 7.51 | 7.81 [7.95] |

**Table 1.** Predicted π-π* gaps for benzene on 2D substrates. The numbers are computed using the point charge model. Provided in square brackets are the numbers computed using the Bader charge model.

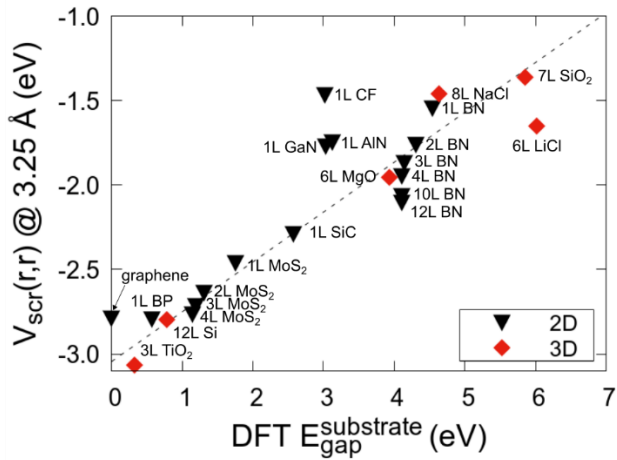

**Figure 4.** Plot of the planar-averaged $V_{scr}(r,r)$ at 3.25 Å away from the substrate against the DFT Kohn-Sham gap $E_{gap}^{substrate}$ of the substrate. Black triangles denote 2D substrates, while red diamonds

denote 3D substrates. A similar plot of planar-averaged $V_{scr}(r,r)$ at 3.25 Å away from the substrate against the GW quasiparticle gap of the substrate is given in Figure S4.

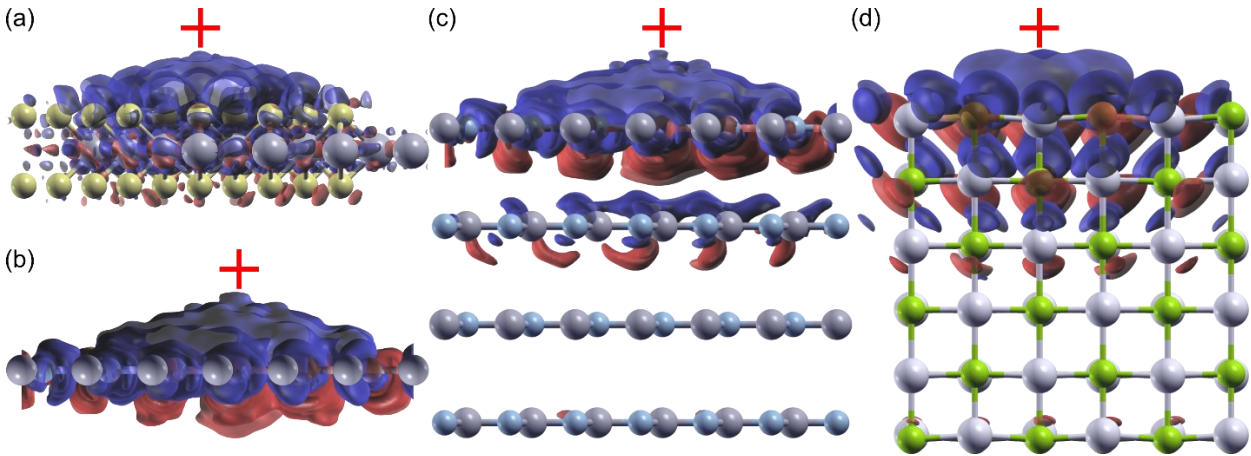

**Figure 5.** The induced charge due to a point electron charge perturbation (red cross) 3.25 Å away from (a-b) thin 2D films (1L MoS$_2$, 1L BN), (b) a thick 2D slab (4L BN), and (c) a 3D slab (LiCl). The same absolute isovalue is used for all plots. Blue denotes depletion of charge, while red denotes accumulation of charge.

As $V_{scr}(r,r)$ evaluated 3.25 Å away from the substrate is a good proxy for the π-π* gap renormalization in adsorbed benzene, we evaluate this quantity for ~20 2D and 3D substrates (the 3D substrates are represented by slabs that are at least 10 Å thick). Both 2D and 3D substrates are chosen to cover a wide range of DFT Kohn-Sham gaps, $E_{gap}^{substrate}$. Interestingly, both 2D and 3D substrates obey the *same* approximately linear relation between $E_{gap}^{substrate}$ and the screening response to a point charge 3.25 Å above the substrate (Figure 4). We note that the value of $V_{scr}$ is rather sensitive to the choice of the surface plane, which, for non-planar surfaces such as CF, LiCl, TiO$_2$, and SiO$_2$, is not obvious. Here we simply choose the surface plane to be the average of the coordinates of the surface atoms, though a more rigorous approach merits investigation. The observed linear behavior is in stark contrast to the well-known fact that screening *within* the 2D material is much weaker than that within 3D materials.[6, 28-30] It also questions the validity of the common practice of computing an effective dielectric constant for a composite system by averaging dielectric constants of individual components[31, 32] – for an atomically thin material, the space occupied by vacuum would be much larger than that by the material, giving an effective dielectric constant close to 1 instead. In fact, using a classical electrostatics model, we can explain why 2D substrates can screen adjacent electronic excitations as well as 3D substrates with similar QP gaps/dielectric constants (see Supplementary Information). Our fully quantum mechanical calculations also show that the induced charge,[21] $\rho_{scr}(r,r')$, which generates the screening response $V_{scr}(r,r')$, is primarily located at the surface (see Supplementary Information; Figure 5). The localization of induced charge at the surface is in agreement with classical electrostatics, which tells us that the polarization-induced charges must lie only at the surface where there is a discontinuity in the dielectric environment. However, the quantum mechanical calculations point to a small distribution of the induced charge above and below the surface plane, with out-of-plane orbitals enabling the atomically thin 2D materials like BN to better screen the perturbing charge. We remark that the linear relation in Figure 4 enables one to estimate the HOMO-LUMO gap of small molecules physisorbed on substrates (2D and 3D), knowing only the gas phase HOMO-LUMO gap and the DFT Kohn-Sham band gap of the substrate. Specifically, this can be done by using the equation for the best-fit line in Figure 4, as follows:

$$E_{gap}^{V_{scr}} = E_{gap}^{gas-phase} - 0.29 E_{substrate}^{gap} + 3.04 \qquad (9)$$

The predicted HOMO-LUMO gaps, shown in Figure 3 (hollow symbols), give remarkably good agreement with the benchmark GW calculations, given the simplicity of equation (9). To enable predictions of HOMO-LUMO gaps for small molecules with different adsorption heights, we provide the corresponding equations of best fit in Table S3 for typical adsorption heights of 3.2-3.6 Å. These results give a more general version of equation (9) that depends on the adsorption height:

$$E_{gap}^{V_{scr}} = E_{gap}^{gas-phase} - (-0.13d + 0.71)E_{substrate}^{gap} + (-0.82d^2 + 7.03d - 17.20) \qquad (10)$$

Using equation (10), we obtain values of $E_{gap}^{V_{scr}}$ within 0.2% of those from equation (9) for $d = 3.25$ Å. Equivalent equations using the GW quasiparticle substrate gap instead of the DFT Kohn-Sham band gap of the substrate, are provided in the Supplementary Information.

In summary, we have used first principles calculations to study dielectric screening by 2D substrates, explicitly computing the electronic screening response at $r$, $V_{scr}(r,r')$ due to a point charge perturbation at $r'$. These calculations enable one to ascertain unambiguously the intrinsic role of 2D materials in screening charged impurities – a task that is not possible experimentally. Our results shed light on the interpretation of recent experiments[10-14] aimed at understanding the effects of screening by 2D substrates, and show that BN layers play an important role in screening out the effect of charged impurities, such as those present on underlying $SiO_2$ substrates. Atomically thin BN monolayers also screen remarkably well, enabled by the out-of-plane orbitals, which allow for the distribution of induced charge to generate a screening response. We also provide simple equations that allow one to estimate rather reliably the HOMO-LUMO gap of small molecules physisorbed on 2D and 3D substrates, using only the gas phase HOMO-LUMO gap, the substrate band gap, and the adsorption height of the molecule. Taken together, our results provide important insights into the nature of 2D screening that will help both experimentalists and theorists to better understand and interpret the non-trivial interactions between 2D materials and adjacent charges.

METHODS

Calculations were performed in BerkeleyGW[33] using wavefunctions and eigenvalues computed in Quantum ESPRESSO[34-36] at the density functional theory (DFT) level.

DFT wavefunctions and eigenvalues were computed within the local density approximation (LDA) for all systems with the exception of those involving $MoS_2$, where the PBE exchange-correlation functional[37] was used in conjunction with optimized norm-conserving Vanderbilt (ONCV) pseudopotentials.[38, 39] The ONCV pseudopotentials allow for the inclusion of Mo semicore electrons without a prohibitively high kinetic energy cutoff for wavefunctions. In $MoS_2$ systems the inclusion of semicore states in the Mo pseudopotential has a negligible effect on the value of $V_{scr}$, however the effect on the QP gap is significant (Table S4). The kinetic energy cutoff for wavefunctions was set to 80 Ry (60 Ry) for LDA (PBE) pseudopotentials.

The static dielectric and self-energy matrices were computed using a 16 Ry cutoff for matrix elements. A 10 Ry cutoff for the sum over bands was imposed for all systems except for the benzene/BN (4x4x1 supercell) interface, which used a 6 Ry cutoff. All GW gaps were computed using the modified static remainder method[40] in order to ensure convergence with respect to the sum over unoccupied bands. The π-π* gaps presented in Table 1 were calculated at the Γ-point.

$V_{scr}(r,r')$ was calculated following equation (1), where $\varepsilon_{G,G'}(q,\omega=0)$ is the static RPA dielectric matrix computed using BerkeleyGW. (We note that our code can handle dielectric matrices produced with or without the non-uniform neck sampling (NNS) method[41] for k-point sampling.) In our implementation, the inverse of the dielectric matrix is multiplied in reciprocal space by the bare Coulomb interaction to obtain $W_{G,G'}(q,\omega=0)$, from which the bare Coulomb interaction is subtracted to give $V_{scr,G,G'}(q,\omega=0)$. We then perform a six-dimensional inverse Fourier transform to get $V_{scr}(r,r';q)$. Finally, we calculate the screening potential as

$$V_{scr}(r,r') = \sum_q e^{iq.(r-r')} V_{scr}(r,r';q) \tag{11}$$

The bare Coulomb interaction used in both the GW and $V_{scr}(r,r')$ calculations was truncated[42] in the direction normal to the interface to avoid spurious interactions between periodic images.

The induced charge,[21] $\rho_{scr}(r,r')$, which generates the screening response $V_{scr}(r,r')$, is given by

$$V_{scr}(r,r') = -\int dr'' \frac{e^2}{|r-r''|} \rho_{scr}(r'',r') \tag{12}$$

$\rho_{scr}(r,r')$ is computed by first constructing the interacting RPA polarizability $\chi_{G,G'}(q,\omega=0)$,[43]

$$\chi_{G,G'}(q,\omega=0) = v_G^{-1}(q) \left( \varepsilon_{G,G'}^{-1}(q,\omega=0) - \delta_{G,G'} \right) \tag{13}$$

Then we have

$$\rho_{scr,G,G'}(q,\omega=0) = \chi_{G,G'}(q,\omega=0) v_{G'}(q) \tag{14}$$

We then perform a six-dimensional inverse Fourier transform to get $\rho_{scr}(r,r';q)$ and obtain

$$\rho_{scr}(r,r') = \sum_q e^{iq.(r-r')} \rho_{scr}(r,r';q) \tag{15}$$

Interfaces were created by placing benzene 3.25 Å atop 3x3x1 substrate supercells (BN, 2L BN, MoS$_2$, and graphene), as well as a 4x4x1 supercell (BN). The Brillouin zones (BZ) of the cells were sampled via a uniform 6x6x1 grid of k-points, with the exception of the 3x3x1 benzene/BN cell, which used an 8x8x1 sampling to ensure commensurability with the corresponding 4x4x1 supercell. In all cases except benzene/2L BN we used the recently-implemented nonuniform neck sampling (NNS) method[41] to ensure convergence with respect to k-point sampling. The omission of NNS in the benzene/2L BN system is justified by tests on BN, which show that the use of NNS has a minimal effect on its band gap and $V_{scr}$ for a 6x6x1 sampling of the BZ. The BZ of the bare substrates were sampled via 18x18x1 k-points and also make use of NNS.

All systems feature 15 Å of vacuum in order to separate periodic images normal to the surface. In separate tests, we verified that increasing the vacuum by a further 10 Å does not alter the value of $V_{scr}$ at 3.25 Å above the surface.

Bader charges were calculated with the Bader Charge Analysis tool[44] using charge densities calculated in VASP.[45]

The 2D autocorrelation, $C(r)$, for a function $f(\vec{r})$, was computed as in Ref. 46:

$$C(r) = \frac{1}{2\pi} \int_0^{2\pi} d\phi \langle\langle f(\vec{r}) f(0) \rangle\rangle, \quad (16)$$

where the angular brackets denote an average over the area and the integration averages over angular orientations. For the autocorrelation of the screened Coulomb potential, we note that the dielectric matrices used to construct $W$ were computed using the unit cell with a Brillouin Zone sampling equivalent to the supercell size in order to prevent artificial periodicity within the cell area. For BN we used a 73x73x1 supercell, corresponding to a charge impurity density of 3.47 x $10^{11}$ cm$^{-2}$. The supercell sizes of MoS$_2$ and graphene, 58x58x1 and 73x73x1, respectively, were chosen to closely match the area, and therefore the impurity density, of the BN supercell.

AUTHOR INFORMATION

**Corresponding Author**

*Email: phyqsy@nus.edu.sg.

**Author Contributions**

§Keian Noori, and Nicholas Lin Quan Cheng contributed equally to this work.

**Notes**

The authors declare no competing financial interest.

ACKNOWLEDGEMENTS

We acknowledge support from Grant NRF-NRFF2013-07 from the National Research Foundation, Singapore and Grant MOE2016-T2-2-132 from the Ministry of Education, Singapore. We thank Miguel Dias Costa for his help in parallelizing codes used in this work and Shaffique Adam for discussions on Ref. 11. Computations were performed on the NUS CA2DM cluster and National Supercomputing Centre Singapore (NSCC) under project IDs: 1100446 and 1100949. We acknowledge support from the Singapore National Research Foundation, Prime Minister's Office, under its medium-sized centre program.

# Supplementary Information for:

# Dielectric Screening by 2D Substrates

*Keian Noori[†,§], Nicholas Lin Quan Cheng[‡,§], Fengyuan Xuan[†], Su Ying Quek[*,†,‡]*

[†]Centre for Advanced 2D Materials, National University of Singapore, 6 Science Drive 2, 117546, Singapore

[‡]Department of Physics, National University of Singapore, 2 Science Drive 3, 117542, Singapore

**Corresponding Author**: *Email: phyqsy@nus.edu.sg.

**Author Contributions:** [§]Keian Noori, and Nicholas Lin Quan Cheng contributed equally to this work.

# Contents

# 1. Classical electrostatic screening of a point charge

For a point charge of charge $q$ at a distance $d$ above the $z$-axis, the electric field is simply given by

$$E_z(z) = \frac{q}{4\pi\epsilon_0} \frac{z-d}{|z-d|^3} \tag{S1}$$

For a point charge of charge $q$ at a distance $d$ above a semi-infinite conductor ($z < 0$), the potential and the electric field (for $z > 0$) is solved by adding an image charge of charge $-q$ on the other side of the plane. The electric field is

$$E_z(z) = \frac{q}{4\pi\epsilon_0} \left( \frac{z-d}{|z-d|^3} - \frac{z+d}{|z+d|^3} \right) \tag{S2}$$

For a point charge of charge $q$ at a distance $d$ above a semi-infinite dielectric ($z < 0$) of dielectric constant $\varepsilon$, the image charge is instead

$$q' = -\frac{\epsilon - 1}{\epsilon + 1} q$$

The corresponding electric field is

$$E_z(z) = \frac{q}{4\pi\epsilon_0} \left( \frac{z-d}{|z-d|^3} - \frac{z+d}{|z+d|^3} * \frac{\epsilon-1}{\epsilon+1} \right) \tag{S3}$$

For a point charge of charge $q$ at a distance $d$ above the top surface ($z = 0$) of a dielectric slab with dielectric constant $\varepsilon$ and thickness $t$, the electric field is given to be[2]

$$E_z(z) = \frac{q}{4\pi\epsilon_0} \left( \frac{z-d}{|z-d|^3} - \frac{z+d}{|z+d|^3} * \frac{\epsilon-1}{\epsilon+1} + \frac{4\epsilon}{(1+\epsilon)^2} \sum_{n=1}^{\infty} \left[ \left(\frac{\epsilon-1}{\epsilon+1}\right)^{(2n-1)} * \frac{z+d+2nt}{|z+d+2nt|^3} \right] \right) \tag{S4}$$

We plot the above 4 quantities in Figure S1, taking $\varepsilon = 5$, $d = 3$ Å and $t = 6$ Å. This models the electric field due to a point charge perturbation 3 Å above a 2D material. Here, the electric field for a thin dielectric slab is almost the same as that of a semi-infinite dielectric with the same dielectric constant, which implies that a thin dielectric slab can screen the electric field due to a point charge perturbation as well as a semi-infinite dielectric with the same dielectric constant. This implies that the similarity in the behavior of 2D and 3D substrates in screening point charge perturbations adjacent to it can be explained using a simple classical model.

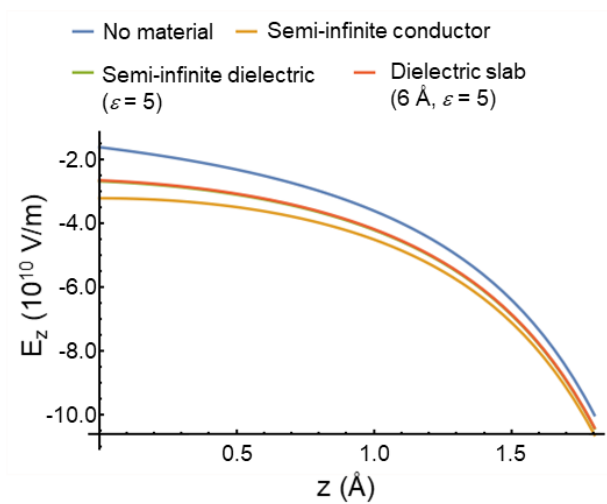

**Figure S1.** Electric field as a function of the distance $z$ from the surface of the material.

### 3. Relation of $V_{scr}$ to the dielectric constant

Previous work on 3D semiconducting substrates demonstrates that there is a clear linear trend between the HOMO-LUMO $\pi$-$\pi^*$ gap of benzene and the band gap of the substrate (computed using LDA).[3] Interestingly, from our results, we find that both 2D and 3D substrates obey the *same* approximately linear relation between the substrate DFT gap and the screening response to a point charge 3.25 Å above the substrate (Figure 4). The dielectric screening capability of a material is most commonly described by its macroscopic dielectric constants, $\varepsilon_{//}$ (out-of-plane) and $\varepsilon_{\perp}$ (in-plane). In Figures S2-3, we show that $V_{scr}(r,r)$ decreases with increasing $\varepsilon_{//}$ and $\varepsilon_{\perp}$, following the same trend irrespective of whether the substrate is 2D or 3D (despite the fact that screening in 2D has strong non-local character[4]).

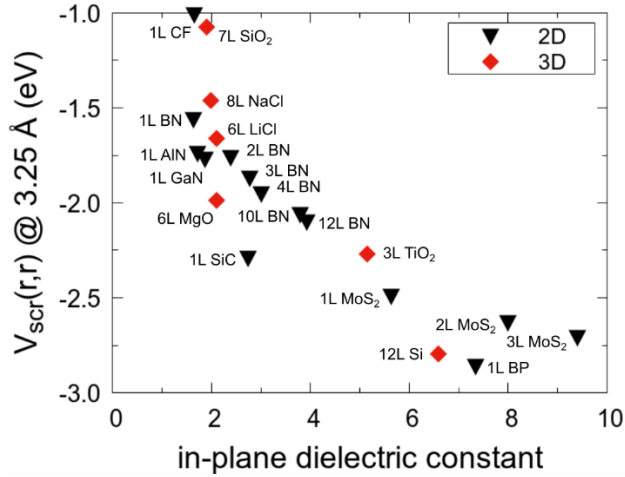

**Figure S2.** Plot of the planar-averaged $V_{scr}(r,r)$ at 3.25 Å away from the substrate against the in-plane dielectric constant of the substrate. Black triangles denote 2D substrates, while red diamonds denote 3D substrates. Computation of the in-plane dielectric constant is performed using density functional perturbation theory (DFPT)[5-6] as implemented in Quantum ESPRESSO[7-8]. The dielectric tensor was computed *ab initio*, where the charge density is computed with a self-consistency threshold for the total energy at $10^{-14}$ Ry, and the DFPT computations are computed with a self-consistency threshold of $10^{-12}$ Ry. and the in-plane dielectric constant is taken to be the diagonal elements of the dielectric tensor along the x and y axes. For black phosphorus (BP), due to the anisotropy of the dielectric tensor along the x and y axes, we take the average of the in-plane dielectric constant over the x and y axes.

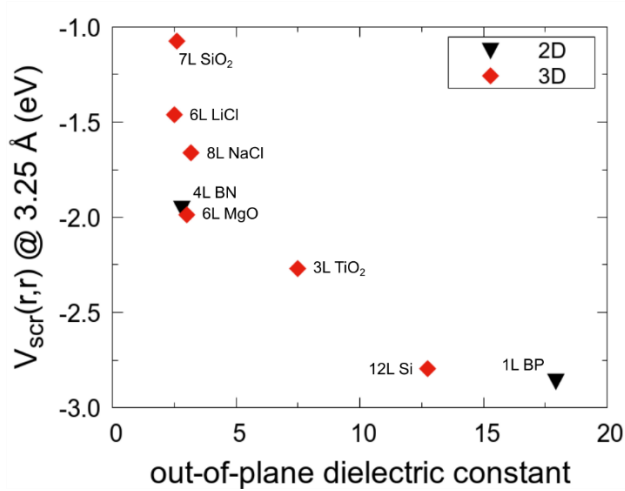

**Figure S3.** Plot of the planar-averaged $V_{scr}(\mathbf{r},\mathbf{r})$ at 3.25 Å away from the substrate against the out-of-plane dielectric constant of the substrate. Black triangles denote 2D substrates, while red diamonds denote 3D substrates. The out-of-plane dielectric constants are computed in DFT following Ref. 9, where the out-of-plane dielectric constant is defined to be the ratio of the changes of the external potential and the total potential due to the application of an external electric field, $\epsilon_\perp = \delta V_{ext}/\delta V$. In our computations, we use an external electric field of magnitude 2.56 V/nm. The value of $\varepsilon_\perp$ is obtained by computing the change in electrostatic potential in response to an applied out-of-plane electric field, however this method does not yield a satisfactory result for atomically thin layered substrates since the resulting change in potential is not necessarily linear.

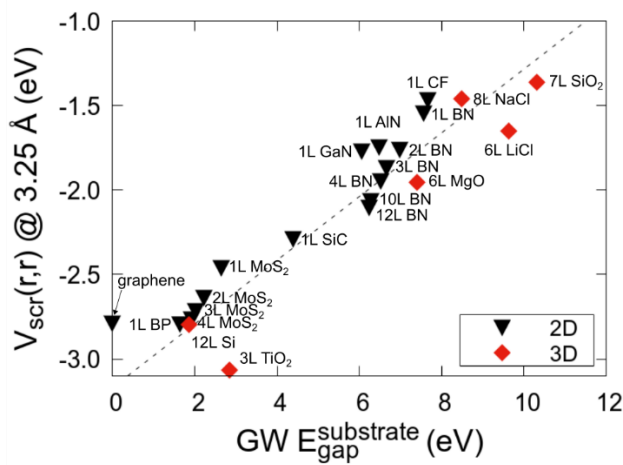

**Figure S4.** Plot of the planar-averaged $V_{scr}(r,r)$ at 3.25 Å away from the substrate against the GW gap $E_{gap}^{substrate}$ of the substrate. Black triangles denote 2D substrates, while red diamonds denote 3D substrates.

## 4. Induced charge due to a point charge perturbation

To understand our results, we explicitly compute the induced charge,[1] $\rho_{scr}(r,r')$, which generates the screening response $V_{scr}(r,r')$, as given by

$$V_{scr}(r,r') = -\int dr'' \frac{e^2}{|r-r''|} \rho_{scr}(r'',r') \quad (S5)$$

The induced charge $\rho_{scr}(r,r')$ represents the charge rearrangement at $r$ in the system due to a point charge perturbation at $r'$. We fix a negative point charge perturbation ~3.25 Å away from the surface and analyze the spatial extent of the induced charge by integrating the induced charge over a cylindrical region of radius $r$ and height $d$ as illustrated in Fig. S5. We maximize the radius of the cylindrical region to $r = 6$ Å and integrate the induced charge over the cylindrical region as a function of $d$.

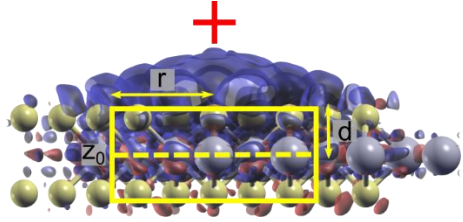

**Figure S5.** Schematic of induced charge integration parameters.

At $d = 4.5$ Å, most of the induced charge is captured in the cylindrical region, which indicated that a large amount of the induced charge is located close to the surface (Fig. S6). Interestingly, the amount of induced charge in the cylinder scales linearly as the planar-averaged $V_{scr}(r,r)$ at 3.25 Å above the surface (Fig. S7). This observation is consistent with plots of the charge response for 2D and 3D surfaces (Fig. 5). Thus, the fact that a 2D material can screen nearby electronic excitations just as well as a 3D material with a similar GW gap may be explained by the observation that screening is predominantly a surface effect. From a classical viewpoint, it is known that the polarization-induced charges must lie only at the surface where there is a discontinuity in the dielectric environment. These quantum mechanical calculations, however, point to a small distribution of the induced charge above and below the surface plane, with out-of-plane orbitals enabling the atomically thin 2D materials like BN to better screen the perturbing charge.

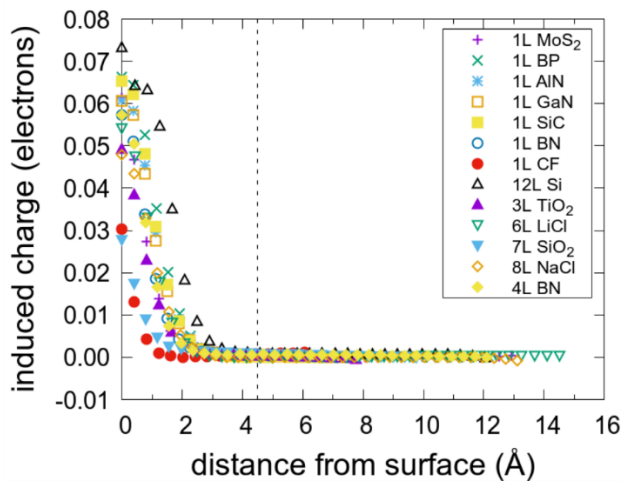

**Figure S6.** Integrated induced charge at $r = 6$ Å as a function of the distance $d$ from the surface atomic plane (see Fig. S4 for definition of $r$ and $d$). The integrated induced charge for all materials is normalized to zero for clarity.

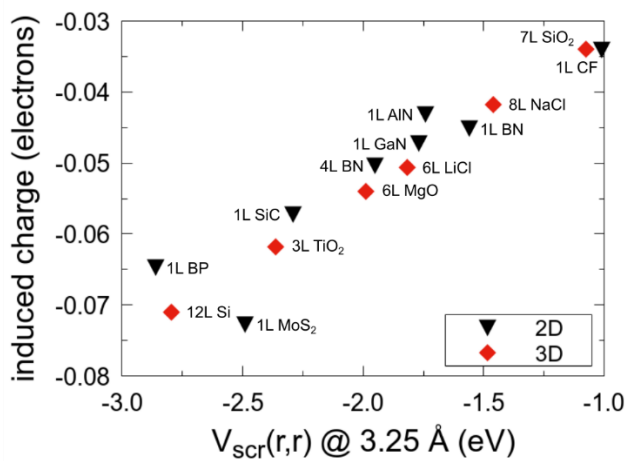

**Figure S7.** Plot of the integrated induced charge against the planar-averaged $V_{scr}(r,r)$ at 3.25 Å away from the substrate. Black triangles denote 2D substrates, while red diamonds denote 3D substrates.

## 5. Tables

| substrate | HOMO projection | LUMO projection |
|---|---|---|
| BN | 0.93 | 0.73 |
| $MoS_2$ | 0.74 | 0.82 |
| graphene | 0.94 | 0.97 |

**Table S1.** Projections of the HOMO and LUMO of isolated monolayer benzene onto the interface states of the interfaces with BN, $MoS_2$ and graphene. The listed values correspond to the relative share of the total projection taken by the interface state most closely resembling the isolated benzene HOMO or LUMO. The total projection for an orbital is normalized to 1.

|  | $d$ containing ~95% of induced charge at $r = 6$ Å (Å) |
|---|---|
| 1L BN | 2.06 |
| 4L BN | 2.54 |
| 1L SiC | 2.25 |
| 1L AlN | 2.44 |
| 1L GaN | 2.25 |
| 1L BP | 2.09 |
| 1L MoS$_2$ | 1.61 |
| 2L MoS$_2$ | 2.03 |
| 3L MoS$_2$ | 2.84 |
| 1L CF | 1.82 |
| 12L Si | 2.47 |
| 7L SiO$_2$ | 1.95 |
| 6L MgO | 1.98 |
| 3L TiO$_2$ | 3.06 |
| 6L LiCl | 2.07 |
| 8L NaCl | 3.09 |

**Table S2.** Height of the cylindrical region, $d$, which contains ~95% of the induced charge with $r = 6$ Å. Schematic of the integrating cylinder in Fig. S5.

| distance from surface (Å) | α [DFT/GW] | β [DFT/GW] |
|---|---|---|
| 3.20 | 0.30 / 0.19 | -3.13 / -3.26 |
| 3.25 | 0.29 / 0.19 | -3.04 / -3.17 |
| 3.30 | 0.29 / 0.18 | -2.96 / -3.08 |
| 3.40 | 0.27 / 0.18 | -2.81 / -2.93 |
| 3.50 | 0.26 / 0.17 | -2.68 / -2.80 |
| 3.60 | 0.27 / 0.16 | -2.56 / -2.67 |

**Table S3.** Parameters for the best-fit line $V_{scr} = \alpha E_{gap}^{substrate} + \beta$ extracted for various heights above the substrate surface for both DFT and GW gaps, $E_{gap}^{substrate}$. Both $V_{scr}$ and $E_{gap}^{substrate}$ are given in eV.

The best-fit equations for $V_{scr}$ using the GW $E_{gap}^{substrate}$ – corresponding to equations (9) and (10) in the main text – are:

$$E_{gap}^{V_{scr}} = E_{gap}^{gas-phase} - 0.19 E_{gap}^{substrate} + 3.17 \tag{S6}$$

$$E_{gap}^{V_{scr}} = E_{gap}^{gas-phase} - (-0.08d + 0.44) E_{gap}^{substrate} + (-0.89d^2 + 7.51d - 18.17) \tag{S7}$$

| system | q→0 treatment | semicore electrons | $V_{scr}$@3.25 Å (eV) | GW gap (eV) |
|---|---|---|---|---|
| 1L MoS$_2$ | MC | yes | -2.11 | 2.68 |
| 2L MoS$_2$ | MC | yes | -2.71 | 2.47 |
| 3L MoS$_2$ | MC | yes | -2.80 | 2.40 |
| 1L MoS$_2$ | MC | no | -2.11 | 2.59 |
| 2L MoS$_2$ | MC | no | -2.71 | 2.09 |
| 3L MoS$_2$ | MC | no | -2.80 | 1.73 |
| 1L MoS$_2$ | NNS | no | -2.49 | 2.52 |
| 2L MoS$_2$ | NNS | no | -2.63 | 2.07 |
| 3L MoS$_2$ | NNS | no | -2.71 | 1.71 |

**Table S4.** Effect of inclusion of semicore electrons and GW self-energy q→0 behavior on GW gap and $V_{scr}$ in MoS$_2$ layers. The treatment of q→0 in the GW self-energy ($\Sigma$) is done via either a Monte Carlo (MC) averaging scheme or the nonuniform neck sampling (NNS) scheme.[20] All pseudopotentials use the LDA exchange-correlation functional.